\documentclass[aps,reprint,article,prx]{revtex4-1}
\usepackage{amsmath}
\usepackage{dcolumn}
\usepackage{graphicx}
\usepackage{float}
\usepackage{multirow}

\begin{document}

\title{Multiband superconductivity and possible nodal gap in RbCr$_{3}$As$_{3}$ revealed by Andreev reflection and single-particle tunneling measurements}

\author{Zhixin Liu,$^1$ Mingyang Chen,$^1$ Ying Xiang,$^1$ Xiaoyu Chen,$^1$ Huan Yang,$^{1,*}$ Tong Liu,$^{2,3,4}$ Qing-Ge Mu,$^{2,3,4}$ Kang Zhao,$^{2,3,4}$ Zhi-An Ren,$^{2,3,4}$ and Hai-Hu Wen$^{1,\ddag}$}

\affiliation{$^1$National Laboratory of Solid State Microstructures and Department of Physics, Collaborative Innovation Center of Advanced Microstructures, Nanjing University, Nanjing 210093, China}
\affiliation{$^2$Beijing National Laboratory for Condensed Matter Physics, Institute of Physics, Chinese Academy of Sciences, Beijing 100190, China}
\affiliation{$^3$School of Physical Sciences, University of Chinese Academy of Sciences, Beijing 100190, China}
\affiliation{$^4$Collaborative Innovation Center of Quantum Matter, Beijing 100190, China}

\begin{abstract}
By measuring point-contact Andreev reflection (PCAR) spectra in the newly discovered chromium-based quasi-one-dimensional superconductor RbCr$_{3}$As$_{3}$, we find clear evidence of two superconducting components, i.e., one having a gap value of about 1.8 meV and another with a gap value of about 5 meV. Since the current injection may have components in both the direction parallel and perpendicular to the [(Cr$_{3}$As$_{3}$)$^{-}$]$_\infty$ chains in the PCAR measurements, it naturally explains the two-component feature observed in this multi-band superconductor. Detailed analysis shows that the larger gap may have an $s$-wave nature. We then carry out the single-particle tunneling measurements based on a scanning tunneling spectroscope by using the needle-like sample as the tip, and in this case the measured current is mainly parallel to the [(Cr$_{3}$As$_{3}$)$^{-}$]$_\infty$ chains. The single-particle tunneling spectra show only one gap feature with a gap value of about 1.8 meV. Fitting to the single particle tunneling spectra indicates that the gap should have a large anisotropy or even node(s). We argue that the absence of the larger gap may be related to the direction of the injecting current. Therefore, our combined experiments show the multi-band superconductivity with one gap being nodal or highly anisotropic.
\end{abstract}

\maketitle
\section{Introduction}
In materials containing 3d transition-metal elements, many new superconductors have been discovered and some of them show the unconventional superconductivity. The latter is argued to have a close relationship with the strong correlation effect of electrons in the 3d orbitals. For the compounds containing 3d transition metal chromium, superconductivity was first observed in the binary CrAs under pressure with a transition temperature $T_c\approx 2$ K \cite{CrAs}. After that some new quasi-one-dimensional (quasi-1D) Cr-based superconductors were discovered at ambient pressure, and they are mainly classified into two families including 233 and 133 with the chemical formulas of $A_2$Cr$_3$As$_3$ and $A$Cr$_3$As$_3$ ($A=$ K, Rb, Cs), respectively \cite{K233,Rb233,Cs233,K133,Rb133}. Recently, the superconductor Na$_2$Cr$_3$As$_3$ was also successfully synthesized \cite{Na233}, and it has the highest $T_c$ = 8.6 K among these Cr-based superconductors. This discovery has been extended to the Mo-based system, superconductivity with $T_c$ of about 10.3 K has been reported in K$_2$Mo$_3$As$_3$ \cite{Mo233Ren}. It has been claimed that the 133 phase may be more stable against water moisture in air than the 233 phase, thus it is easier to be handled for many elegant measurements. The structural characteristic of the Cr-based materials is that it contains the 1D [Cr$_{3}$As$_{3}$]$_\infty$ chains separated by alkali metal atoms \cite{K233}, which makes the physical properties and superconductivity of these materials very interesting.

Clearly, the newly discovered Cr-based superconductors provide a new platform to research on the superconductivity related to the pairing of 3d electrons. The band structure calculations of $A_2$Cr$_3$As$_3$ reveal three sets of Fermi surfaces (FSs) which are mainly contributed by 3d orbitals of chromium. The FSs consist of two flat ones from quasi-1D $\alpha$ and $\beta$-bands as well as a three-dimensional (3D) one from 3D $\gamma$-band \cite{CaoCCal,DaiJHPRL,HuJPTripletPRB,ZhouYCal}. The quasi-1D FSs were then confirmed by angle-resolved photoemission spectroscopy measurements in K$_2$Cr$_3$As$_3$ \cite{ARPES}. Because of the complex band structure, different kinds of unconventional pairing symmetries were predicted for $A_2$Cr$_3$As$_3$ superconductors theoretically. One interesting proposal is that a nodal triplet $p_z$-wave pairing driven by the ferromagnetic fluctuations within the Cr sublattice occurs on the quasi-1D $\beta$ band \cite{HuJPTripletPRB}. Another prediction is that the pairing can change from a spin-triplet state $f_{y(3x^2-y^2)}$ with line nodes on the 3D $\gamma$-band to a spin-triplet fully gapped state $p_z\hat{z}$ at the quasi-1D $\alpha$-band when the strength ratio between Hund's coupling and inter-orbital repulsion decreases \cite{ZhouYCal}. Meanwhile from some other theoretical works, the quasi-1D and 3D FSs may couple with each other, which results in a strongly anisotropic singlet pairing and gap nodal rings on the 3D Fermi sphere when the local repulsion is strong enough \cite{KimCal}.

Up to now many experimental results suggest unconventional superconductivity in $A_2$Cr$_3$As$_3$. The very large upper critical field $H_{c2}$ exceeding the Pauli limit may suggest a strong and unconventional pairing mechanism \cite{CaoZhuReview,K233,Rb233,BalakirevHc2}. Furthermore, in K$_2$Cr$_3$As$_3$, it was found that $T_c$ decreases significantly with increase of the density of non-magnetic impurities, which may suggest the sign-reversal gaps of superconductivity \cite{CaoGHImpurity}. The similar conclusions have also been obtained from the absence of the Hebel-Slichter coherence peak of $1/T_1$ just below $T_c$ in NMR studies \cite{ImaiNMR,ZhengGQNMR}. In addition, $1/T_1$ decreases rapidly below $T_c$ following a function of $T^5$ at low temperatures in Rb$_2$Cr$_3$As$_3$, which may be a proof of the presence of point nodes in the gap function \cite{ZhengGQNMR}. The possible line-nodal gap was proved by low-temperature electronic specific-heat data \cite{LuoJLSH} and the muon-spin relaxation data \cite{uSR} in K$_2$Cr$_3$As$_3$, as well as the penetration depth measurements in K$_2$Cr$_3$As$_3$ and Rb$_2$Cr$_3$As$_3$ \cite{YuanHQPD1}. The anisotropic superconductivity was observed in $A_2$Cr$_3$As$_3$ not only when the magnetic field is applied along or perpendicular to the $c$-axis, but also with the in-plane rotation of field \cite{CaoZhuReview}, and the threefold in-plane $H_{c2}$ modulation was observed and argued to be related to the spin-triplet pairing \cite{ZhuZWInplanAnis}.

The $A$Cr$_3$As$_3$ family has half amount of the alkali metal atoms comparing to the $A_2$Cr$_3$As$_3$ family, and the space groups for these two families are different \cite{K133,Rb133}. However, they both contain the 1D [Cr$_{3}$As$_{3}$]$_\infty$ chains but with different ionic valences of the Cr$_{3}$As$_{3}$ units. The normal-state electrical resistivity also has different temperature dependent behaviors in these two families of the materials \cite{K133,Rb133}. Theoretical calculation predicted that FSs are constructed by two or three quasi-1D and two 3D bands \cite{DaiJHCal,YangFCal}. The superconducting pairing symmetry can be triplet $f$ or $p_z$ wave or singlet $s^\pm$ wave with different interaction parameters \cite{YangFCal}. The nodal superconducting gap structure in RbCr$_3$As$_3$ was inferred from the thermal conductivity measurements which show a sizable value of the thermal conductivity coefficient in the zero temperature limit at ambient field, and the magnetic field enhancement is clearly faster than that expected for a full-gap superconductor \cite{LiSYTC}. However, more experiments are required to derive the information of the superconducting gap in the Cr-based family.

Andreev reflection and single-particle tunneling measurements are very useful and combinatorial to detect the superconducting gap. In this work, we will present a detailed study of the superconducting gap structure on the RbCr$_3$As$_3$ single crystals by using the point-contact Andreev reflection (PCAR) and scanning tunneling microscopy/spectroscopy (STM/STS) measurements. The results show the existence of two superconducting components. The large gap has the value of about 5 meV, and shows only on the Andreev reflection spectra; while the smaller one with the value of about 1.8 meV can be observed both on the Andreev reflection and the STS measurements. We suggest that the smaller gap is highly anisotropic or even nodal based on the fittings to the STS data.

\section{Experimental methods and superconductivity characterization}

\begin{figure}
\includegraphics[width=8cm]{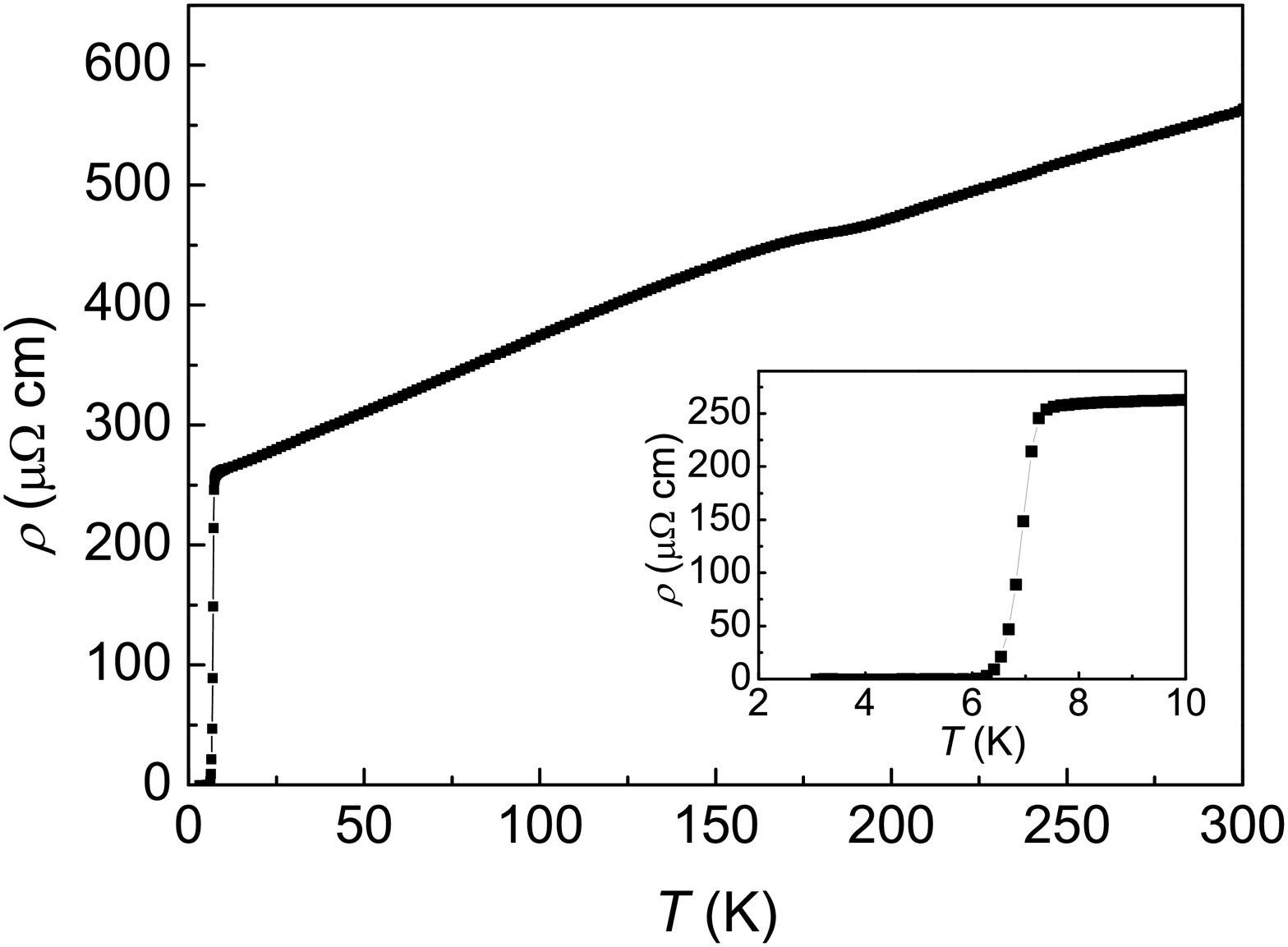}
\caption{Temperature dependence of resistivity of a separate needle-like RbCr$_3$As$_3$ single crystal. The inset shows the enlarged view of the $\rho$-$T$ curve near the superconducting transition.} \label{fig1}
\end{figure}

The RbCr$_3$As$_3$ single crystals were grown by the deintercalation process of Rb$^+$ ions from Rb$_2$Cr$_3$As$_3$ precursors \cite{Rb133}. The samples are loosely assembled by many needle-like single crystals with diameter of sub- to several micrometers, and the $c$-axis is defined as the direction along the needle main axis or the [(Cr$_{3}$As$_{3}$)$^{-}$]$_\infty$-chains of the sample. The RbCr$_3$As$_3$ samples are stable in the air. The X-ray diffraction data measured on a bundle of crystals can be found in the previous work \cite{Rb133} measured by a Bruker single-crystal X-ray diffractometer, and all the diffraction peaks can be well indexed with the hexagonal space group P6$_3$/m.

The needle-like single crystals are loosely assembled to form a bundle, and there is large separation between these needles; hence a direct measurement of the cross-sectional size of a bundle-like sample induces a very large error for determining the resistivity. In order to make a more elegant resistivity measurement, we separate a needle-like single crystal from a bundle of samples. The electrical resistivity was measured in a physical property measurement system (PPMS-9, Quantum Design) by using the standard four-probe method, and the current is applied along the $c$-axis of the sample. The widths and the thicknesses of the samples are detected by scanning electron microscope (SEM, Phenom ProX), and the typical values are from 3 to 7 $\mu$m for these samples. Figure~\ref{fig1} shows temperature dependent resistivity for a separate RbCr$_3$As$_3$ single crystal at 0 T, and the inset shows the enlarged view near the superconducting transition.The onset transition temperature $T_{c}$ is 7.17 K determined from the 90\% of the normal-state resistivity, and the transition width is about 0.64 K determined by using criterion of 10\% and 90\% of the normal-state resistivity. The residual resistance ratio $RRR=\rho(300\ \mathrm{K})/\rho(8\ \mathrm{K}>T_c)=2.17$ is consistent with previous reports \cite{Rb133,LiSYTC}, but the measured residual resistivity $\rho(8\ \mathrm{K})=249$ $\mu\Omega\cdot$cm here is much smaller than the values obtained previously \cite{Rb133}. The resistivity in our measurements are reproducible, and the difference is about 20\% for different samples. The obtained resistivity value is even smaller than that obtained by weighing the mass to calculate the cross-section \cite{LiSYTC}. This may be caused by the disconnections located on some needle-like crystals, which increases the total resistivity in that work. We note that the resistivity here is also much larger than the values of $A_2$Cr$_3$As$_3$, and the possible explanation for this may be the crystal defects and lattice deformation induced by the deintercalation process \cite{Rb133,K133}. In addition, temperature dependent behaviors of resistivity are very different between these two systems \cite{Rb133,K133}, and the resistivity increases almost linearly with increase of temperature in RbCr$_3$As$_3$. This feature is similar to the one observed in cuprates which have a typical non-Fermi-liquid behavior \cite{resistivity}, and it may be related to the spin scattering due to the magnetic frustration \cite{DaiJHCal} in the material. The much larger resistivity in $A$Cr$_3$As$_3$ than that in $A_2$Cr$_3$As$_3$ also needs further investigation.

The samples for PCAR or STS measurements are also divided from a bundle of single crystals, and the typical sizes of selected samples are 300-500 $\mu$m in length and about several micrometers in width. For the PCAR measurements, the point contact was constructed by sticking a thin gold wire with the diameter of 16 $\mu$m to the sample by using a tiny bit of silver paste (DuPont, 4929N), and a typical configuration is shown in the inset of Fig.~\ref{fig2}(a). The length of the sample in the point-contact region was about 20 $\mu$m, and two ends of the sample were buried in silver paste more than 100 $\mu$m in length to reduce the contact resistance of these electrodes. Since the contact constructed by silver paste has a finite size, the current injection may have components both parallel and perpendicular to the [(Cr$_{3}$As$_{3}$)$^{-}$]$_\infty$-chains. The PCAR measurements were carried out in PPMS-9 by using a home-made setup. The ac modulation for the differential conductance measurement is several microamperes with the frequency of 985 Hz in the lock-in measurements.

The STS measurements were carried out in an ultra-high vacuum, low-temperature and high-magnetic field scanning tunneling microscope (USM-1300, Unisoku Co., Ltd.). We made the STS tip by a separate RbCr$_3$As$_3$ single crystal in the following steps. We first found a very thin wire of the sample by tweezers, and then cut off one end to get a relatively fresh terminal for the STS tip. Afterwards, we used the silver paste to glue most part of the sample on a Pt/Ir tip, and made the newly-cut prominent end face towards a flat Au(111) single-crystal flake sample (purity 99.999\%). The single-particle tunneling spectra were recorded between the needle-like sample tip and the Au single crystal in an ultra-high vacuum chamber with a base pressure of about $1\times10^{-10}$ torr. The direction of injecting current is mainly along the [(Cr$_{3}$As$_{3}$)$^{-}$]$_\infty$-chains or $c$-axis of the sample for the STS measurements. The setpoint condition for the spectrum measurements is $I_{set}=50$ pA and $V_{set}=10$ mV, and the ac modulation is 0.5 mV with the frequency of 871.773 Hz.

\section{Point-contact Andreev reflection results}
\begin{figure}
\includegraphics[width=8cm]{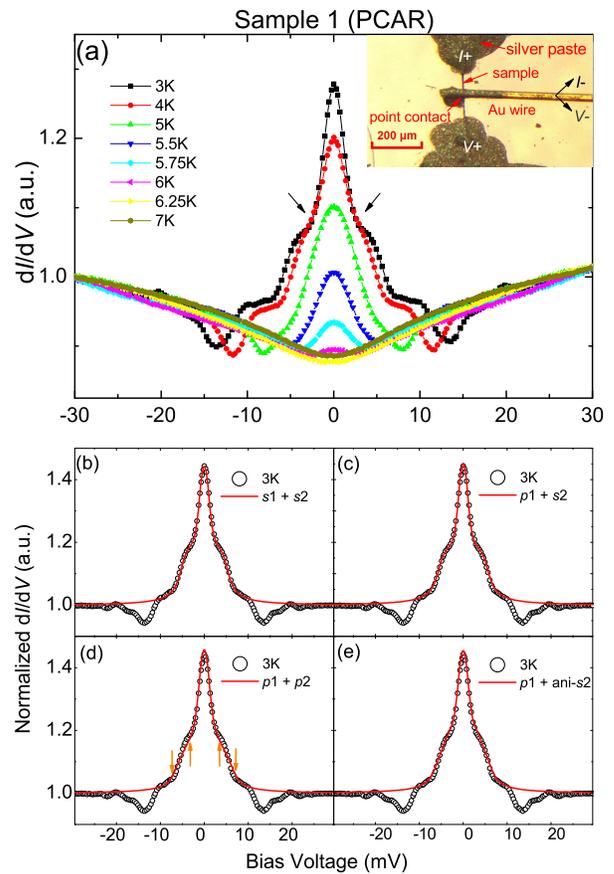}
\caption{(a) Raw data of PCAR spectra measured at various temperatures in RbCr$_3$As$_3$ Sample-1. The inset shows the photograph of the point-contact junction for the Andreev reflection measurement. The point contact is constructed by a little bit of silver paste attaching to a gold wire with the diameter of 16 $\mu$m, and two ends of the sample are buried in the silver paste in order to lower down the contacting resistance. The junction resistance $R_j =159\ \Omega$ at 3 K. (b-e) Normalized PCAR spectra (open circles) at 3 K divided by the one at 7 K, and the BTK fitting results (the solid lines) with different kinds of double superconducting gaps. For the smaller gap, the related fitting parameters are $\Delta_{s1}=1$ meV and $\Gamma_{s1}=0.43$ meV for the $s$-wave gap, or $\Delta_{p1}=1.4\sin\theta$ meV, $\Gamma_{p1}=0.39$ meV for the $p$-wave gap; for the larger gap, the related fitting parameters are $\Delta_{s2}=5$ meV and $\Gamma_{s2}=0.65$ meV for the $s$-wave gap, $\Delta_{p2}=6.8\sin\theta$ meV, $\Gamma_{p2}=0.4$ meV for the $p$-wave gap, or $\Delta_{\mathrm{ani-}s2}=6.4(0.2\cos2\theta+0.8)$ meV and $\Gamma_{\mathrm{ani-}s2}=0.4$ meV for the anisotropic $s$-wave gap. The other fitting parameters are $w_1=0.75$ and $Z_1=Z_2=0.1$ for the two gaps. }\label{fig2}
\end{figure}

Figure~\ref{fig2}(a) shows the measured PCAR spectra at different temperatures and 0 T. Since the upper critical field is very high, e.g., $\mu_0H_{c2}=72.4$ T in the zero-temperature limit and $\mu_0H_{c2}\approx52$ T at 3 K for RbCr$_3$As$_3$ samples \cite{Rb133}, the highest magnetic field of 9 T provided by our PPMS gives only negligible influence on the spectrum at 3 K. For this reason, we only show the temperature dependence of PCAR spectra to investigate the superconducting gap of this material. One can see that there are always zero-bias peaks on the spectra taken below $T_c$. This kind of peak often appears on the spectra taken by the PCAR measurements on an unconventional superconductor, especially when the injecting current is along the nodal direction. In cuprate superconductors, this has been well investigated both in theory \cite{PCAR cuprate theory1,PCAR cuprate theory2,PCAR cuprate theory3} and by experiments \cite{LauraGreene,WeiJ,PCAR cuprate experiment}. It should be noted that there are a pair of hump features at about $\pm5$ meV on the spectrum measured at 3 K, and they are marked by black arrows in Fig.~\ref{fig2}(a). We will show that this corresponds to the second superconducting gap in RbCr$_3$As$_3$ through the fitting. Beside the superconducting gap features, some unexpected small dips of $dI/dV$ can be observed at low temperatures and energies just above the superconducting gap energy, e.g., from 10 to 15 meV for the spectrum measured at 3 K. The bottom energies of these dips decrease with increase of temperature. This kind of dip feature seems to be very common in point-contact measurements and was argued to be induced by the critical current effect or the heating effect when the voltage was high and the point contact was not in the pure ballistic limit \cite{JcPC}. From temperature dependence of the spectra, one can see that the height of the zero-bias peak decreases with increase of temperature, and the spectrum shape changes to a V-shaped normal-state background when the temperature reaches 7 K $\approx T_c$. To get rid of the normal-state background signal, we normalized the spectrum measured at 3 K by dividing the one measured at 7 K, and the normalized curve is shown by the open circles in Fig.~\ref{fig2}(b-e). The differential conductance at zero-bias is about 1.45 times larger than the value of the high-energy normal-state background for the normalized spectrum at 3 K, which suggests that the barrier height of the junction is small and the contact is closer to the Andreev reflection (normal metal/superconductor, NS) limit instead of the tunneling (normal metal/insulator/superconductor, NIS) limit.

\begin{figure}
\includegraphics[width=8cm]{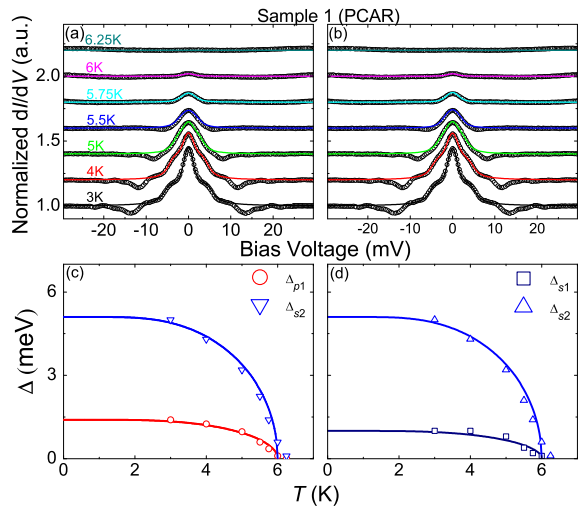}
\caption{(a,b) The normalized PCAR spectra (open circles) divided by the spectrum measured at 7 K, and the two-band BTK fitting results (solid lines) by using (a) a $p$-wave smaller gap and an isotropic-$s$-wave larger gap and (b) an isotropic-$s$-wave smaller gap and an isotropic-$s$-wave larger gap. For two different kinds of double gap models, the fitting parameters $w_1=0.75$ and $Z_1=Z_2=0.1$ all keep constant for all the spectra at different temperatures, while $\Gamma_1$ and $\Gamma_2$ change slightly with increase of temperature during the fitting procedure. (c,d) Temperature dependence of the superconducting gaps from fitting results (open symbols) by BTK model with (c) a $p$-wave smaller gap and an isotropic-$s$-wave larger gap, and (d) two isotropic-$s$-wave gaps. The solid lines are the theoretical curves from the weak-coupling BCS model.} \label{fig3}
\end{figure}

In order to get in-depth understanding of superconducting gaps from the PCAR spectra, we fit the data with the Blonder-Tinkham-Klapwijk (BTK) theory \cite{BTK}. The expression of differential conductance $G=dI/dV$ from BTK theory reads,
\begin{equation}
G(V)=\int_{-\infty}^{+\infty}\frac{df(E-V,T)}{dV}\left[1+A(E)-B(E)\right]dE,
\end{equation}
where $f(E,T)$ is the Fermi distribution function, and $A(E)$ and $B(E)$ represent the contributions from the Andreev reflection and normal reflection, respectively. Then the PCAR spectra can be described by the superconducting gap $\Delta$, the junction barrier height $Z$, the broadening factor $\Gamma$, and temperature $T$ \cite{PCAR cuprate theory3}.
Since there is an obvious two-gap feature on the PCAR spectrum measured at 3 K, we use a two-gap BTK model \cite{MgB2PCAR} to fit the normalized differential conductance $G$ as
\begin{equation}
G=w_1G_1+(1-w_1)G_2.
\end{equation}
Here $w_1$ is the weight for the smaller gap, and $G_1$ and $G_2$ are the normalized differential conductances originated from different bands with smaller and larger superconducting gaps, respectively. The BTK fitting results by different kinds of superconducting gaps are shown in Fig.~\ref{fig2}(b-e). It is found that the PCAR spectra can be well fitted by the BTK model. For the $p$-wave gap, we use the gap function of $\Delta_p \sin\theta$. The gap minimum $\Delta_\mathrm{min}=-\Delta_p$ while the gap maximum $\Delta_\mathrm{max}=+\Delta_p$, and then the ratio $\Delta_\mathrm{min}/\Delta_\mathrm{max}=-1$. For the anisotropic-$s$-wave gap, we use the gap function of $\Delta_{\mathrm{ani-}s}(x\cos2\theta+1-x)$, and $\Delta_\mathrm{min}/\Delta_\mathrm{max}=1-2x$ which reflects the gap anisotropy. One can see that the fitting results are very consistent with the measured spectra whether the smaller gap takes an $s$-wave [Fig.~\ref{fig2}(b)] or a $p$-wave form [Fig.~\ref{fig2}(c)]. Therefore, we can not make judgement whether the smaller gap has the gap nodes from the fitting results. However, for the larger gap, the $p$-wave gap function cannot fit the data well with any fitting parameters, and an example is shown in Fig.~\ref{fig2}(d). The fitting curve deviates from the experimental data in the ranges marked by the orange arrows. We then use the anisotropic-$s$-wave gap function to fit the larger gap, and the fitting requires the gap anisotropy $x\leq0.2$ or $\Delta_\mathrm{min}/\Delta_\mathrm{max}\geq0.6$. The fitting result with $x=0.2$ is shown in Fig.~\ref{fig2}(e). Hence for the larger gap, there may be a little gap anisotropy exists, but the possibility for the nodal gap is rare from the fitting results. In order to obtain the temperature dependent behavior of the superconducting gap, we use the two-gap BTK model with different gap functions to fit the normalized PCAR spectra at different temperatures, and the fitting results are shown in Fig.~\ref{fig3}(a,d). Figure~\ref{fig3}(c,d) show the obtained temperature dependence of the two superconducting gaps derived from the fittings, while the solid lines are the theoretical calculations carried out by using the Bardeen-Cooper-Schrieffer (BCS) gap model.

\begin{figure}
\includegraphics[width=8cm]{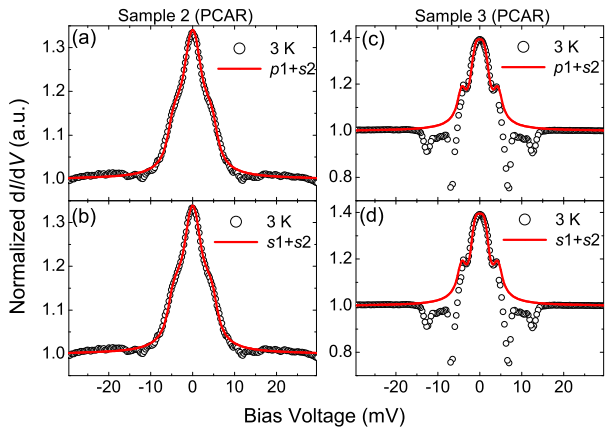}
\caption{(a,b) The normalized PCAR spectrum (open circles) for Sample-2 ($R_j =17\ \Omega$) and the two-gap BTK fitting results (solid lines) with (a) $p$+$s$-wave gaps and (b) two $s$-wave gaps at 3 K. The fitting parameters for $p$+$s$-wave gaps are $\Delta_{p1}=1.8\sin\theta$ meV, $\Delta_{s2}=5.3$ meV, $w_1=0.75$, $\Gamma_{p1}=0.9$ meV, $\Gamma_{s2}=0.7$ meV, and $Z_{p1}=Z_{s2}=0.1$. The fitting parameters for two $s$-waves gaps are $\Delta_{s1}=1.3$ meV, $\Delta_{s2}=5.3$ meV, $w_1=0.75$, $\Gamma_{p1}=0.95$ meV, $\Gamma_{s2}=0.7$ meV, and $Z_{s1}=Z_{s2}=0.1$. (c,d) The normalized PCAR spectrum for Sample-3 ($R_j =4\ \Omega$) and the BTK fitting curves with (c) $p$+$s$-wave gaps and (d) two $s$-wave gaps at 3 K. The junction resistance is about 4 $\Omega$. The fitting parameters for $p$+$s$-wave gaps are $\Delta_{p1}=2.2\sin\theta$ meV, $\Delta_{s2}=4.5$ meV, $w_1=0.75$, $\Gamma_{p1}=0.32$ meV, $\Gamma_{s2}=0.35$ meV, $Z_{p1}=0.28$, and $Z_{s2}=0.75$. The fitting parameters for two $s$-wave gaps are $\Delta_{s1}=1.7$ meV, $\Delta_{s2}=4.5$ meV, $w_1=0.75$, $\Gamma_{p1}=0.39$ meV, $\Gamma_{s2}=0.35$ meV, $Z_{p1}=0.2$, and $Z_{s2}=0.75$.} \label{fig4}
\end{figure}

In order to verify our experimental results, we carried out the PCAR experiments on different samples, and the data for two other samples are shown in Fig.~\ref{fig4}. The normalized spectra measured in Sample-2 and Sample-3 show the similar two-gap feature as in Sample-1, but with slightly different gap values obtained from the fitting procedures. The spectrum weight from the band(s) with smaller gap is much larger than that with the larger gap, and the weight ratios are all about 3:1.

BTK theory can only describe the spectra measured on the contacts in the ballistic regime, and we thus check the status of the contact junction. The residual residual resistivity $\rho(8\ \mathrm{K})\approx259\ \mu\Omega\cdot$cm in RbCr$_{3}$As$_{3}$. If we take a simplified model of a spherical Fermi surface with average Fermi momentum $\overline{k_F}=0.5\pi/a_0$ \cite{DaiJHCal,YangFCal} and the lattice constant $a_0=9.37$ \AA \  \cite{Rb133}, we can estimate the mean free path $l\approx17$ nm. The junction resistance varies from 159 to 4 $\Omega$ in our point-contact measurements, which corresponds to the size of the contact $a$ from 11 to 68 nm based on the Sharvin formula $R_s=4\rho l/(3\pi a^2)$ \cite{PCreview}. Here the mean free path approximately equals to the contact size, which means that the contact may be between the diffusive regime and the ballistic regime. Although the fitting formulas are different in these two regimes, the obtained superconducting gap values are very close to each other \cite{PCreview}. In addition, the point contact is usually consisted by many individual contacts which have bigger contact resistances and smaller sizes \cite{PCreview}. So the contact here may still locate in the ballistic regime.

We have two more arguments to address the reason that the point-contact junction is not in the thermal regime. Firstly, the differential conductivity increases with increasing of the bias voltage in high bias-voltage range as shown in Fig.~\ref{fig2}(a), which means that the junction resistance decreases with increasing of the bias voltage. This contradicts what expected in the thermal regime of a point contact, since the contact resistance would increase due to stronger heating effect with larger bias voltage. In addition, the V-shaped spectra obtained above $T_c$ by PCAR measurements are similar to the ones measured by STS. Secondly, the barrier height $Z$ from the fitting procedure is always very small for all three point-contact junctions, which means that the contact is close to the ballistic regime or the intermediate regime. From the reasons mentioned above, we conclude that the contact is not in the thermal region, and we can get the correct gap information from the measured PCAR spectra.

\section{Scanning tunneling spectroscopy results}

\begin{figure}
\includegraphics[width=8cm]{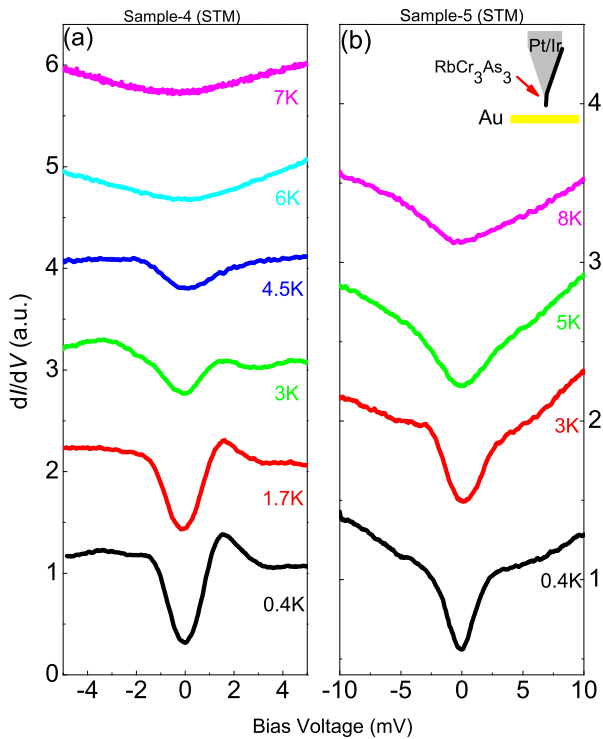}
\caption{Temperature dependent tunneling spectra measured by scanning tunneling microscopy at different temperatures for (a) Sample-4 and (b) Sample-5. The inset in (b) displays the schematic configuration of STS measurements, and the tunneling spectra are taken between the needle-like sample of RbCr$_3$As$_3$ attached to the Pt/Ir tip and a flat gold flake with the junction resistance of about 0.2 G$\Omega$. } \label{fig5}
\end{figure}

\begin{figure}
\includegraphics[width=8cm]{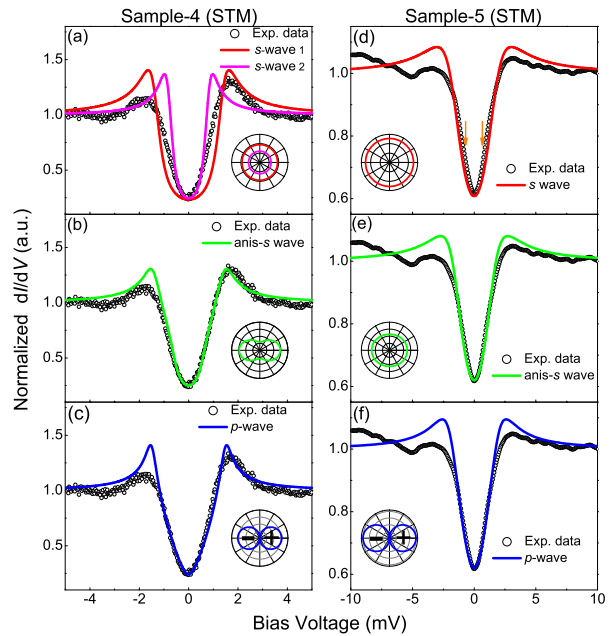}
\caption{STS spectra (open circles) at 400 mK normalized by the one measured at 7 K for Sample-4 (a-c) and at 8 K for Sample-5 (d-f). The solid lines in (a-f) show different fitting curves by Dynes model and with different gap functions for Sample-4 and Sample-5. The superconducting gap functions used for the fitting are plotted in the insets of (a-f). The fitting parameters are (a) $\Delta_{s1}=1.26$ meV, and $\Gamma_{s1}=0.35$ meV; $\Delta_{s2}=0.8$ meV, and $\Gamma_{s2}=0.2$ meV; (b) $\Delta_{ani-s}=1.46(0.3\cos2\theta+0.7)$ meV, and $\Gamma_{ani-s}=0.24$ meV; (c) $\Delta_p=1.56\sin\theta$ meV, and $\Gamma_{p}=0.15$ meV; (d) $\Delta_s=1.7$ meV, and $\Gamma_{s}=1.3$ meV; (e) $\Delta_{ani-s}=1.58(0.05\cos2\theta+0.95)$ meV, and $\Gamma_{ani-s}=1.18$ meV; (f) $\Delta_p=1.97\sin\theta$ meV, and $\Gamma_{p}=0.91$ meV.} \label{fig6}
\end{figure}

The above-mentioned Andreev reflection results show two-superconducting-gap feature clearly. However, we can not judge whether the smaller gap is nodal or nodeless from the PCAR data and fitting results presented above. We then do the further measurements by STS at temperature as low as 400 mK. The needle-like sample, which was attached to the end of the Pt/Ir tip, was used as the tip for the STS measurements. The tunneling spectra were taken between the sample and a flat gold flake as illustrated by the schematic figure in the inset of Fig.~\ref{fig5}(b). The tunneling spectra measured at different temperatures are displayed in Fig.~\ref{fig5}(a) and (b) for Sample-4 and Sample-5, respectively. One can see that the normal state spectra taken above $T_c$ show the V-shape, which is somewhat similar to the normal-state background obtained by PCAR measurements. In addition, the superconducting spectra are of the typical single-particle tunneling type with only one pair of coherence peaks for both samples, which means that we detect mainly the contribution from one superconducting component. We use the spectra measured above $T_c$ as the backgrounds and show the normalized spectra at 400 mK in Fig.~\ref{fig6}(a-f) for Sample-4 and 5. Then the Dynes model \cite{Dynes}, which can be regarded as the situation in BTK model with $Z=\infty$, was used to fit the normalized spectra using different kinds of gap functions. The fitting results are also shown in Fig.~\ref{fig6}(a-f) as solid lines. For Sample-4, the zero-bias differential conductance is as low as about 24\% of the normal state background far above the superconducting gap. Obviously from Fig.~\ref{fig6}(a), one can see that the fitting curves by using an isotropic $s$-wave gap with any fitting parameters cannot catch up the main features of the measured spectrum. This suggests that the gap anisotropy is necessary for the superconducting gap from the fitting procedure. We find that the minimum anisotropy for the fitting corresponds to the ratio of $\Delta_\mathrm{min}/\Delta_\mathrm{max}=0.4$, and the fitting result is shown in Fig.~\ref{fig6}(b). In addition, the fitting by a $p$-wave gap can also catch up the main features of the experimental data as shown in Fig.~\ref{fig6}(c) for Sample-4. For Sample-5, the zero-bias differential conductance is about 61\% of the normal state value, and the in-gap suppression of density of states is much smaller than that in Sample-4. Nevertheless, the model by an isotropic $s$-wave gap cannot fit the experimental data well in the energy ranges near the arrow positions. The gap anisotropy is also necessary for the fitting procedure. The minimum gap anisotropy expressed by $\Delta_\mathrm{min}/\Delta_\mathrm{max}=0.9$ is required for the fitting, and the fitting result is shown in Fig.~\ref{fig6}(e). Besides, the model with a $p$-wave gap can also fit the experimental data well as shown in Fig.~\ref{fig6}(f).

The STS data show the clear existence of the superconducting gap with the gap value from 1.46 to 1.97 meV which is consistent with the smaller gap values from the PCAR measurements. However, the normalized zero-bias conductance ranging from about 0.24 to 0.6 makes it difficult to determine the precise gap function of this gap. The finite zero-bias conductance may be attributed to the impurity scattering effect for a nodal superconductor. Of course the finite zero-bias conductance can also be induced by the imperfection of the tip and a large scattering effect is inevitable. Although the origin for the zero-bias conductance is unclear yet, we confidently conclude that the gap anisotropy is necessary for the fitting procedure.

\section{Discussion and conclusion}

\begin{figure}
\includegraphics[width=8.5cm]{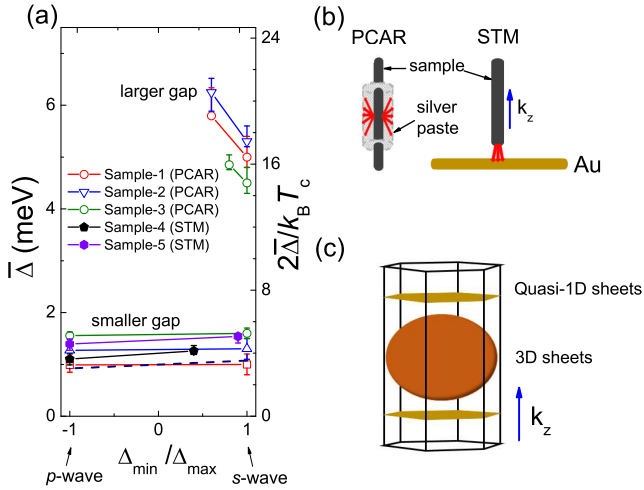}
\caption{(a) Average superconducting gap values $\overline{\Delta}$ and their corresponding gap ratios ${2\overline{\Delta}}/{k_BT_c}$ obtained from different samples by the fittings based on different gap functions. The horizontal coordinate $\Delta_\mathrm{min}/\Delta_\mathrm{max}$ reflects the anisotropy of the gap function. The PCAR spectra measured in Sample 1-3 are fitted by two-component BTK model, and the STS spectra measured in Sample 4 and 5 are fitted by single-gap Dynes model. The error bars of the gap values are determined by the fitting procedure with slight change of other fitting parameters. The solid lines are used to schematically connect two fitting parameter and show the range of the gap anisotropy in which the experimental data can be well fitted by the Dynes model. The dashed line connects the values ${2\overline{\Delta}}/{k_BT_c}=3.53$ for an $s$-wave gap and 3.03 for a $p$-wave gap predicted by weak-coupling BCS theory.  (b) Schematic image of the current injecting directions (red arrows) for PCAR and STS measurements. (c) Schematic image of the simplified Fermi surface of $A$Cr$_{3}$As$_{3}$, which consists of two sets of three flat sheets from the quasi-1D bands and two 3D sheets.} \label{fig7}
\end{figure}

In order to determine the coupling strength, we need to evaluate the gap ratio $2\Delta/k_BT_c$. For the anisotropic or nodal gap, it is better to use the average gap value to calculate the gap ratio. Thus we use the formula $\overline{\Delta}^{2}=\frac{1}{2\pi}\int_0^{2\pi}\Delta^{2}(\theta)d\theta$ to calculate the average gap $\overline{\Delta}$. All the obtained average gap values are plotted in Fig.~\ref{fig7}(a) from different fittings to the spectra measured in different samples. The average gap values change with gap functions or gap anisotropy ratios $\Delta_\mathrm{min}/\Delta_\mathrm{max}$. In Fig.~\ref{fig7}(a), the solid lines, which just connects two data points, are used for guidance; they cover the range of anisotropy $\Delta_\mathrm{min}/\Delta_\mathrm{max}$ in which the experimental data can be well fitted based on the corresponding gap function.
The broadening factor $\Gamma$ used for the fitting in Sample-4 is smallest among all the values for the fittings in other samples, and the fitting provides a constrain that the smaller gap is highly anisotropic or even nodal. If the smaller gap takes a nodal $p$-wave form as predicted by the theory, the mean value of the average gap is $\langle\overline{\Delta}_{p1}\rangle=1.26\pm0.23$ meV, and the calculated gap ratio is ${2\langle\overline{\Delta}_{p1}}\rangle/{k_BT_c}=4.1\pm0.8$. Such gap ratio is larger than 3.03 from weak coupling BCS theory for a $p$-wave gap. If the gap is an anisotropic $s$-wave one with $\Delta_\mathrm{min}/\Delta_\mathrm{max}=0.4$, the average gap value is just a little bit larger than that from the $p$-wave gap fitting. For the larger gap obtained from the PCAR spectra, it should be nodeless from the fitting procedure and the largest gap anisotropy corresponds to a minimum $\Delta_\mathrm{min}/\Delta_\mathrm{max}$ value of about 0.6. If the larger gap is an isotropic $s$-wave gap, the mean value of the gap is $\langle{\Delta}_{s2}\rangle=5.0\pm0.43$ meV which corresponds to a gap ratio of ${2\langle{\Delta}_{s2}\rangle}/{k_BT_c}\approx16$. This gap ratio is much larger than 3.53 from weak coupling BCS theory for an $s$-wave gap. The large values of the gap ratio suggest the strong coupling and unconventional superconductivity in this material, and could suggest a pairing probably not mediated by phonons. We note that the recent specific heat measurements revealed a single nodal gap in single crystal of K$_2$Cr$_3$As$_3$; the corresponding  ${2\overline{\Delta}}/{k_BT_c}\approx11.1$ \cite{LuoJLSH} which is also a very large value and may be consistent with our observation of the larger gap in RbCr$_3$As$_3$. Another penetration depth measurement reported a possible $d$-wave gap with ${2\overline{\Delta}}/{k_BT_c}\approx3.1$ in polycrystalline K$_2$Cr$_3$As$_3$ \cite{YuanHQPD}, and such ratio is similar to the smaller gap ratio from our measurements in RbCr$_3$As$_3$. In this point of view, there may also be two-gap feature in K$_2$Cr$_3$As$_3$, while the two gaps were selectively detected by different measurement techniques. This is similar to our experimental result here.

From the PCAR data, the weight of the larger gap $w_2$ is about $(25\pm5)\%$ from the fitting procedure, but this weight seems to be negligible for the STS spectra. The possible reason for this result may be related to the different configurations for two kinds of measurements. As shown in Fig.~\ref{fig7}(b), the current injection takes some stochastic directions in the PCAR measurements, while the tunneling current is mainly along the $k_z$ direction for the STS measurements. $A$Cr$_3$As$_3$ is a typical multi-band system, and a simplified schematic image of the Fermi surface is shown in Fig.~\ref{fig7}(c). The Fermi surface consists of three flat sheets from quasi-1D bands and two 3D sheets from 3D bands \cite{DaiJHCal,YangFCal}. The density of states contributed by the 3D sheets is much larger than those from the flat sheets \cite{DaiJHCal,YangFCal}. In the STS measurements, we see only one component with the gap value comparable to the smaller one derived in the PCAR experiments. In the STM measurements the tunneling matrix element problem may weaken the detection of the 3D band(s) \cite{STMreview1,STMreview2}. Here in RbCr$_3$As$_3$, the 3D Fermi surfaces are composed by the Cr-$3d_{x^2-y^2}$ and $3d_{xy}$ orbitals, and the 1D Fermi surfaces have the contribution from Cr-$3d_{z^2}$ orbital according to the theoretical calculations \cite{DaiJHCal,YangFCal}. Therefore it is possible that the Au single crystal with an approximate $s$-wave electronic structure have much weaker wave function overlapping with the planar Cr-$3d_{x^2-y^2/ xy}$ orbital bands \cite{DavisMatrix,Zhenyu}, which leads to a negligible tunneling matrix element along $k_z$ direction from the 3D Fermi surface of RbCr$_3$As$_3$. Consequently, the major contribution of the tunneling spectra may come from the 1D Cr-$3d_{z^2}$ orbits in STS measurements. However, the reason why we cannot detect the larger superconducting gap by STS measurements needs further theoretical explanations. Since our measurements do not have the capability to assign superconducting gaps to FSs in momentum space, we can only conclude that there is a multi-gap feature with strong-coupling superconductivity in RbCr$_3$As$_3$. In addition, the smaller gap, which was observed by both PCAR and STS measurements with different configurations, is anisotropic or even nodal; while the larger gap is nodelss. According to the theoretical predictions, the superconducting pairing symmetry can be triplet $f$-, $p_z$- or singlet $s^\pm$-wave with different interaction parameters \cite{YangFCal}. Our data and fitting results provide useful information for future theoretical work.

In summary, we observe clear two-gap feature from PCAR measurements in RbCr$_3$As$_3$. The gap values yielded from the fitting to the PCAR spectra are about 1.8 and 5 meV. The larger gap may be nodeless. In the STS measurement by using the needle-like sample RbCr$_3$As$_3$ as the tip, we see however only one gap with the value comparable with that of the smaller gap yielded from the PCAR measurements. The smaller gap may be highly anisotropic or even with gap node(s) inferred from STS measurements and related fitting. The absence of the larger gap in the STS measurements may be explained as that the injecting current is mainly along the $k_z$ direction. The large ratio of $2\overline{\Delta}/k_BT_c$ suggests the unconventional superconductivity in this material.

\begin{acknowledgments}
We appreciate the kind help in the PCAR measurements by Cong Ren, Lei Shan and Xingyuan Hou, and useful discussions with Yi Zhou, Xin Lu and Zhenyu Wang. This work was supported by the National Key R\&D Program of China (Grant No. 2016YFA0300401), the National Natural Science Foundation of China (Grant No. 11534005, and No. 11774402), and the Strategic Priority Research Program of Chinese Academy of Sciences (Grant No. XDB25000000).
\end{acknowledgments}

$^*$ huanyang@nju.edu.cn

$^\dag$ hhwen@nju.edu.cn

\end{document}